\documentclass[10pt]{article}
\usepackage{amsmath}
\usepackage{amssymb}
\usepackage{graphicx}
\usepackage{cite}
\usepackage[labelfont=bf,labelsep=period,justification=raggedright]{caption}

\makeatletter
\renewcommand{\@biblabel}[1]{\quad#1.}
\makeatother
\date{}

\begin{document}

\begin{flushleft}
{\Large
\textbf{Different Reactions to Adverse Neighborhoods in Games of Cooperation}
}\sffamily
\\[3mm]
\textbf{Chunyan Zhang,$^{1,2}$ Jianlei Zhang,$^{1,2}$ Franz J.
Weissing,$^{2,\dag}$ Matja{\v z} Perc,$^{3,\S}$ \\Guangming
Xie,$^{1,\ast}$ Long Wang$^{1}$}
\\[2mm]
{\bf 1} State Key Laboratory for Turbulence and Complex Systems, College of Engineering, Peking University, Beijing, China
{\bf 2} Theoretical Biology Group, University of Groningen, The Netherlands
{\bf 3} Department of Physics, Faculty of Natural Sciences and Mathematics, University of Maribor, Slovenia
\\[1mm]
$^{\ast}$xiegming@pku.edu.cn\\
$^{\S}$matjaz.perc@uni-mb.si\\
$^{\dag}$f.j.weissing@rug.nl
\end{flushleft}
\sffamily

\section*{Abstract}
In social dilemmas, cooperation among randomly interacting
individuals is often difficult to achieve. The situation changes if
interactions take place in a network where the network structure
jointly evolves with the behavioral strategies of the interacting
individuals. In particular, cooperation can be stabilized if
individuals tend to cut interaction links when facing adverse
neighborhoods. Here we consider two different types of
reaction to adverse neighborhoods, and all possible mixtures between
these reactions. When faced with a gloomy outlook, players can
either choose to cut and rewire some of their links to other
individuals, or they can migrate to another location and establish
new links in the new local neighborhood. We find that in general
local rewiring is more favorable for the evolution of cooperation
than emigration from adverse neighborhoods. Rewiring helps to
maintain the diversity in the degree distribution of players and
favors the spontaneous emergence of cooperative clusters. Both
properties are known to favor the evolution of cooperation on
networks. Interestingly, a mixture of migration and rewiring is even
more favorable for the evolution of cooperation than rewiring on its
own. While most models only consider a single type of reaction to
adverse neighborhoods, the coexistence of several such reactions may
actually be an optimal setting for the evolution of cooperation.

\section*{Introduction}
Cooperation is a fascinating area of research since it touches upon
so many different disciplines, ranging from biology to economics,
sociology, and even theology \cite{nowak_11}. The fact that in human
societies cooperative behavior is common among unrelated people is
puzzling from evolutionary point of view, since cooperation can
easily be exploited by selfish strategies. Evolutionary game theory
\cite{maynard_82, hofbauer_98, Weissing_10} provides a theoretical
framework to address the subtleties of cooperation among selfish
individuals. In particular, the prisoner's dilemma
\cite{axelrod_84,roca_plr09} is a paradigm example for studying the
emergence of cooperation in spite of the fact that self-interest
seems to dictate defective behavior.

Past research has identified several key mechanisms (comprehensively
reviewed in \cite{nowak_s06}) that promote the evolution of
cooperation. In particular, spatial reciprocity \cite{nowak_n92b}
has launched a spree of activity aimed at disentangling the role of
the spatial structure by the evolution of cooperation. The seminal
works in this area focused on regular graphs and lattices
\cite{nowak_n92b, nowak_ijbc93, lindgren_pd94, durrett_tpb94,
grim_jtb95, szabo_pre98, brauchli_jtb99, hauert_c03, hauert_n04,
szolnoki_epjb09, szolnoki_njp08}. Later attention shifted to more
complex networks \cite{albert_rmp02, boccaletti_pr06}, and, in
particular, to scale-free networks. Evolutionary games on graphs and
networks are thoroughly reviewed in \cite{szabo_pr07}. More recent
studies have elaborated on various aspects, including the dynamical
organization \cite{gomez-gardenes_prl07}, clustering
\cite{assenza_pre08} and mixing patterns \cite{rong_pre07,
poncela_pre11}, as well as memory \cite{wang_wx_pre06}, robustness
\cite{poncela_njp07}, phase transitions \cite{floria_pre09} and
payoff normalization \cite{santos_jeb06, szolnoki_pa08}.

While considering population structure is an important step for
understanding the evolution of cooperation, a crucial ingredient is
still missing. In real social networks the interaction structure is
frequently not static but evolving in concern with the behavior of
the interacting agents. As reviewed in \cite{perc_bs10} a vibrant
new research area is emerging that studies the joint evolution of
interaction structure and behavior. Many models have focused on the
way players make (or break) links in reaction to the degree of
cooperation they experienced from their interaction
partners\cite{zimmermann_pre04, eguiluz_ajs05, pacheco_jtb06,
pacheco_prl06, santos_ploscb06, hanaki_ms07, szolnoki_epl08,
pacheco_jtb08, kun_a_bs09, szolnoki_epl09, tanimoto_pa09,
van-segbroeck_prl09, graeser_epl09, szolnoki_njp09, lazaro_pre11}. Other models
have considered the possibility to leave uncooperative neighborhoods
\cite{majeski_c99, vainstein_pre01, vainstein_jtb07, helbing_acs08,
helbing_pnas09, droz_epjb09, meloni_pre09, roca_pnas11}. It is
plausible that both mechanisms can promote the evolution of
cooperation, but it is not obvious which of the two mechanisms is
more efficient. Moreover, it is not self-evident that all
individuals use the same rules for changing their interaction
network in response to adverse conditions. In fact, everyday
experience tells us that different people may react quite
differently when they find themselves in a bad neighborhood: while
some tend to migrate to another location, others tend to stay put
and instead search for new friends (or get rid of old friends) in
order to improve the situation.

Motivated by such considerations, we study the joint
evolution of cooperation and interaction structure in the prisoner's
dilemma game and in the snowdrift game. The players are of two types
that differ in the way they react to an adverse neighborhood. A
fixed fraction $m$ of the players consists of `migrants', who in
proportion to the number of defectors in their neighborhood tend to
migrate to another (unoccupied) position in the network. The
complementary fraction $1-m$ of the players consists of `rewirers',
who have the tendency to break their links with defectors and
subsequently to reattach the free links to other players. By
changing the parameter $m$, our model allows to transverse smoothly
from an adaptive linking model ($m=0$) to a migratory model ($m=1$).
In between these two extremes, we have a situation where different
players react differently when finding themselves in an adverse
neighborhood.

\section*{Results}

Our analysis is based on the prisoner's dilemma game and the
snowdrift game, two paradigm models for the evolution of
cooperation. We label the payoff parameters in line with the
conventions for the prisoner's dilemma \cite{axelrod_84}: a
cooperating player receives the ``reward'' $R$ in case of mutual
cooperation and the ``sucker's payoff'' $S$ in case of being
defected; a defecting player receives the ``temptation to defect''
$T$ when the other player cooperates and the ``punishment'' $P$ in
case of mutual defection. By definition, a prisoner's dilemma game
satisfies the payoff relationships $T>R>P>S$. When played as a
one-shot game in a well-mixed population, defect is the only
evolutionarily stable strategy; despite of the fact that the payoff
$P$ to both players can be considerably smaller than the payoff $R$
for mutual cooperation. The snowdrift game is characterized by $T >
R$ and $S>P$. When played as a one-shot game in a well-mixed
population, none of the two pure strategies is evolutionarily stable
and a mixed strategy is expected to result \cite{hauert_n04}.
Without loss of generality, we normalize $R$ and $S$ to $R=1$ and
$S=0$. In all our graphs, $T$ is systematically varied
from $1$ to $2$. Hence, the game considered is a prisoner's dilemma
game if $P>0$ and a snowdrift game if $P<0$.

\begin{figure}
\begin{center}\includegraphics[width=11cm]{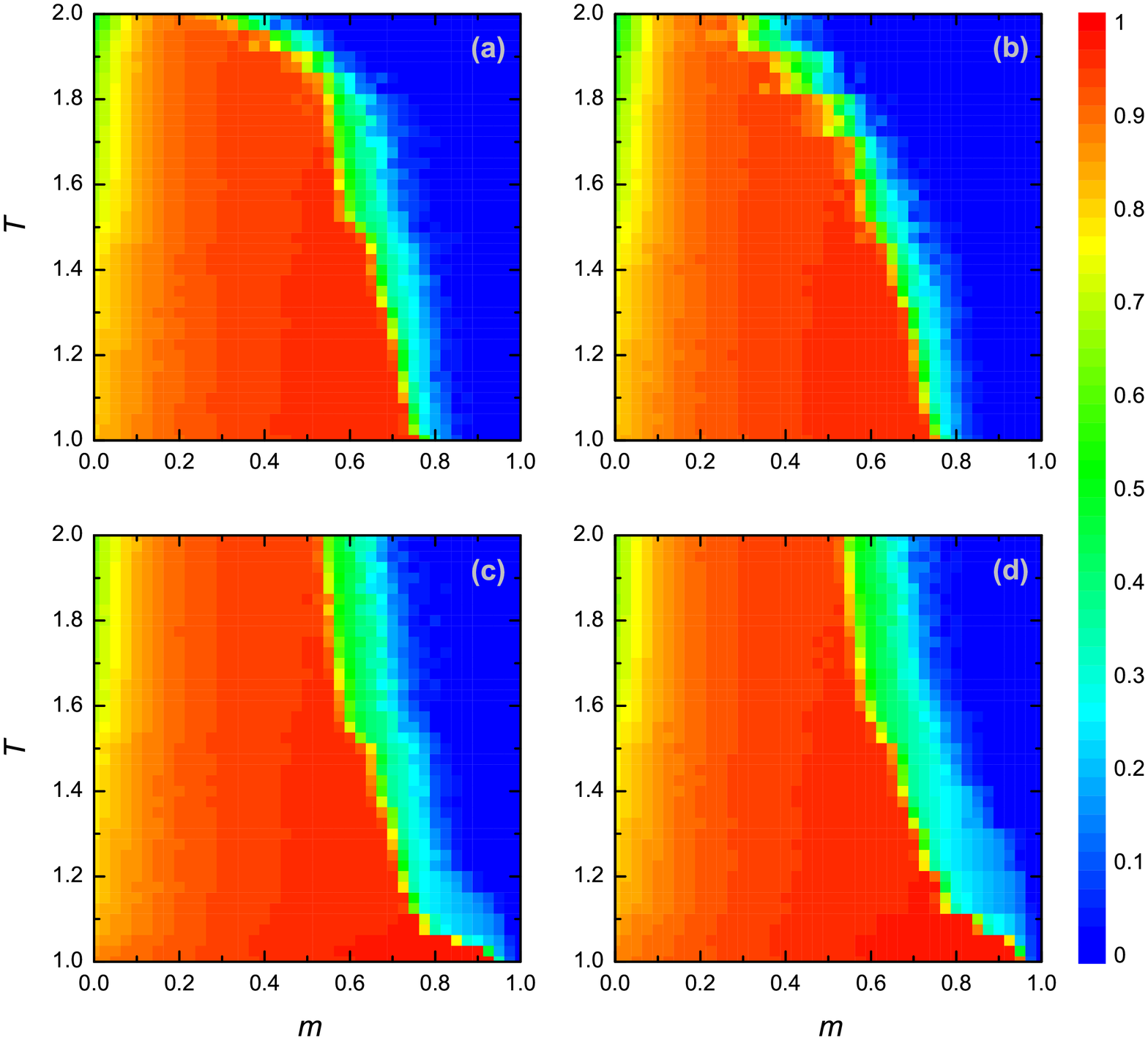}\end{center}
\caption{\textbf{Stationary density of cooperators in dependence on the fraction of players reacting to adverse conditions by migration ($m$) and the payoff parameter ($T$).} Panels (a) and (b) depict the outcome for the prisoner's dilemma game with $P=0.05$ and $P=0.10$, respectively, while panels (c) and (d) depict the outcome for the snowdrift game with $P=-0.05$ and $P=-0.10$, respectively. Other payoff parameters are $R=1$ and $S=0$ in all four panels. The density of cooperators in the stationary state (after $10^6$ iteration steps) is color-coded from blue (full defection) to red (full cooperation), as indicated on the right of the figure. If all players are ``migrants'' ($m=1$; right-hand border of each panel), defection is clearly dominant, while intermediate levels of cooperation evolve if all players are ``rewirers'' ($m=0$; left-hand border of each panel). In all types of game, the highest level of cooperation
evolves in mixed populations consisting of both migrants and rewirers. In this figure, the density of occupied nodes is $\rho=0.8$.}
\label{fig1}
\end{figure}

Fig.~\ref{fig1} shows how for four values of the payoff
parameter $P$ the level of cooperation evolves in relation to the
temptation $T$ to defect and the relative frequency $m$ of players
reacting to adverse conditions by migration. By and large, the
outcome is very similar in all four cases. Cooperation is more
difficult to achieve for larger values of $T$, but in general the
outcome is dominated by the parameter $m$. If migration is the only
reaction to adverse conditions ($m=1$; right-hand border of each
panel), cooperation goes extinct not only in the prisoner's dilemma
games (upper panels in Fig.~\ref{fig1}) but also in the snowdrift
games (bottom panels in Fig.~\ref{fig1}). In contrast, cooperation
can reach high levels or even go to fixation if all players react to
adverse conditions by rewiring ($m=0$; left-hand border of each
panel). Interestingly, a combination of migration and rewiring is
most favorable for the evolution of cooperation. For a broad range
of $m$-values ($0.1<m<0.6$), cooperation tends to fixation, even in
case of a relatively large temptation $T$ to defect. Apparently,
cooperation is favored if a certain fraction of the players choose
for a complete reset of their interactions when surrounded by
defectors, while too high levels of migration mix up the population
to such an extent that local structures providing a foothold for
cooperation cannot develop.

To further analyze the mechanisms enhancing or impeding
cooperation, we performed extensive numerical simulations for the
case $P=0$. This is a border case that is sometimes called a
``weak'' prisoner's dilemma game \cite{roca_plr09}. In general, the
weak form of the prisoner's dilemma game can have other properties
than the strong form \cite{Ohtsuki06}, but Fig.~\ref{fig1} clearly
demonstrates that this is not the case in our model.

\begin{figure}
\begin{center}\includegraphics[width=16cm]{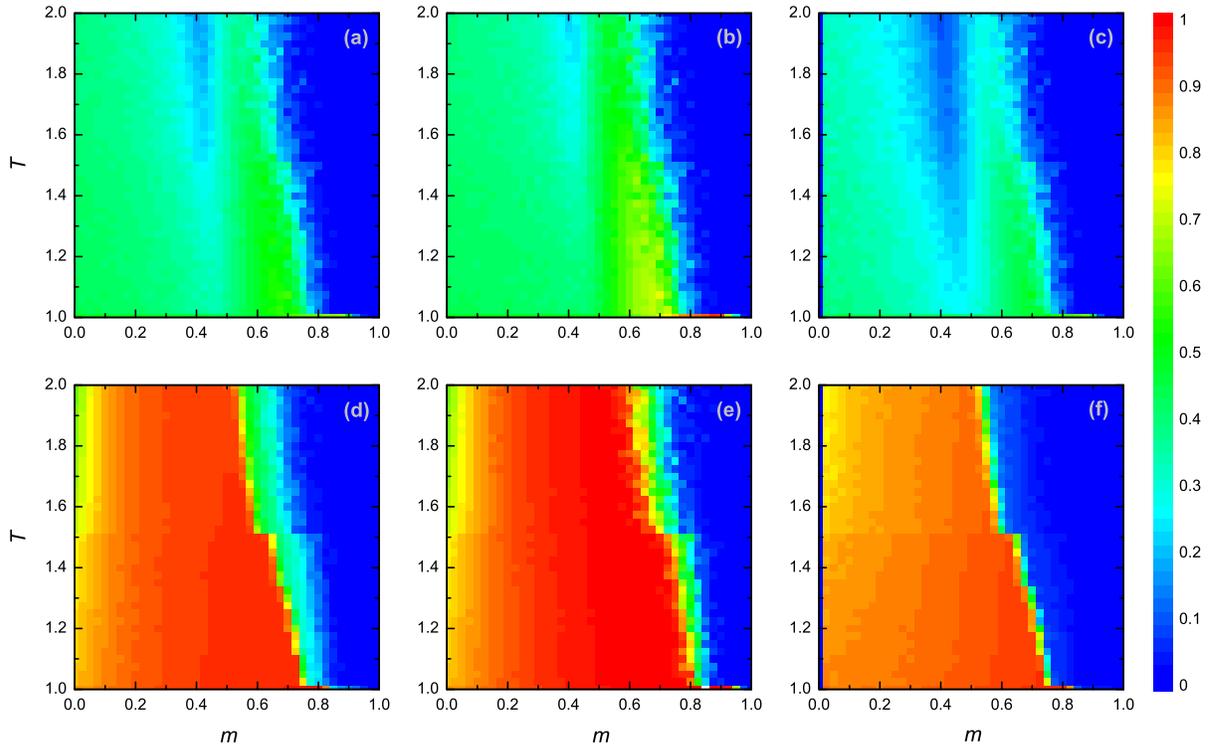}\end{center}
\caption{\textbf{Stationary density of cooperators amongst all players (a,d), amongst players adopting rewiring (b,e), and amongst players adopting migration (c,f) for two different population densities [$\rho=0.4$ in (a,b,c) and $\rho=0.8$ in (d,e,f)].} As
in Fig.~\ref{fig1}, each panel shows the outcome in dependence on the fraction of players reacting to adverse conditions by migration ($m$) and the payoff parameter ($T$), color coded as indicated on the right of the figure. Other payoff parameters are $R=1$ and $S=P=0$ in all six panels. See main text for further details.}
\label{fig2}
\end{figure}

Fig.~\ref{fig2} shows how the evolution of cooperation is affected
by population density. At low densities ($\rho=0.4$; upper panels),
cooperation does not get off the ground and at best stays at the
initial level. If the majority of the population reacts to adverse
conditions by migration, cooperation goes extinct. Similar results
were obtained when the value of $\rho$ was smaller than about $0.5$.
If the population density is too small, players only have few
interaction partners, making it difficult for cooperators to form
local clusters enforcing their success.

Up to now, we have focused on the behavior of the system as a whole.
We will now zoom in a bit and study the different types of player in
more detail. The middle and right panels of Fig.~\ref{fig2}
demonstrate that in the stationary state the strategy choice
(cooperate versus defect) of a player becomes associated with the
player's reaction to adverse conditions. In fact, players adopting
rewiring (Fig.~\ref{fig2}(b)(e)) show a markedly higher tendency to
cooperate than players adopting migration (Fig.~\ref{fig2}(c)(f)).
The difference in cooperation tendency between both types of player
is smallest at low population densities (here $\rho=0.4$), where for
neither type of player the relative frequency of cooperation exceeds
$0.5$. One reason for this may be that the migration of players
provides an opportunity for defectors to invade and destroy the
sparse cooperative clusters in the scattered population, essentially
creating a situation comparable to well-mixed conditions. In the
low-density scenario, cooperation is only sustained (at intermediate
level) when the exploitation from defectors is not too strong. At
high densities, the situation is markedly different. Even for
relatively large values of $m$, there is a boost of cooperation even
for large values of $T$. Still, players adopting migration do worse
than those adopting rewiring.

\begin{figure}
\begin{center}\includegraphics[width=11.015cm]{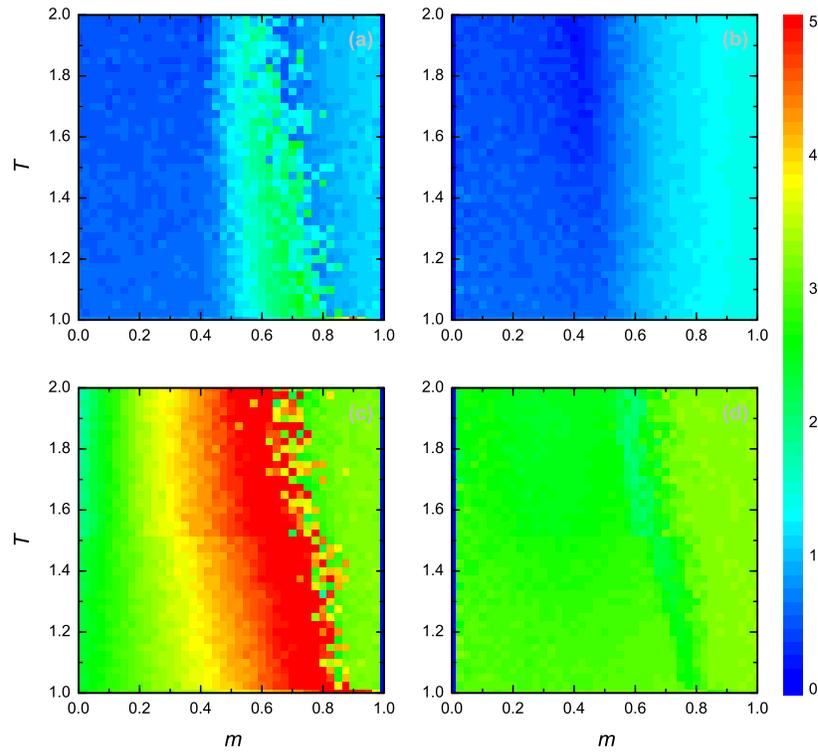}\end{center}
\caption{\textbf{Average degree of players adopting rewiring (a,c) and players adopting migration (b,d) for two different population densities [$\rho=0.4$ in (a,b) and $\rho=0.8$ in (c,d)].} The average initial degree of each occupied node was $4 \rho$. Graphical conventions and payoff parameters are the same in Fig.~\ref{fig2}, only that the color bar encodes the average degree in the stationary state. See main text for further details.}
\label{fig3}
\end{figure}

Fig.~\ref{fig3} shows how the reaction of a player to adverse
conditions affects the player's degree of connectedness in the
evolved stationary population. It is obvious that the topology of a
player's neighborhood is at least partly shaped by the player's
behavioral choices. It has been shown that the adaptive interplay
between the players' strategies and the underlying network can lead
to the emergence of heterogeneity from an initially homogeneous
connectivity structure \cite{nowak_n92b}. As before, the difference
in connectivity between players adopting migration and players
adopting rewiring is quite small at low population density (upper
panels in Fig.~\ref{fig3}). This difference becomes much more
pronounced at high density. For players adopting rewiring (left
panels in Fig.~\ref{fig3}), the average degree first increases with
$m$ (to reach highest levels for $m=0.6$), subsequently decreasing
with a further increase of $m$. In contrast, the average degree of
players adopting migration only marginally depends on $m$, staying
close to the initial value of $4 \rho$. For most parameter
combinations, players adopting rewiring have a higher degree than
players adopting migration. Presumably, rewiring leads to an
accumulation of links between cooperators and thereby the formation
of tightly connected cooperative clusters.

\begin{figure}
\begin{center}\includegraphics[width=11.015cm]{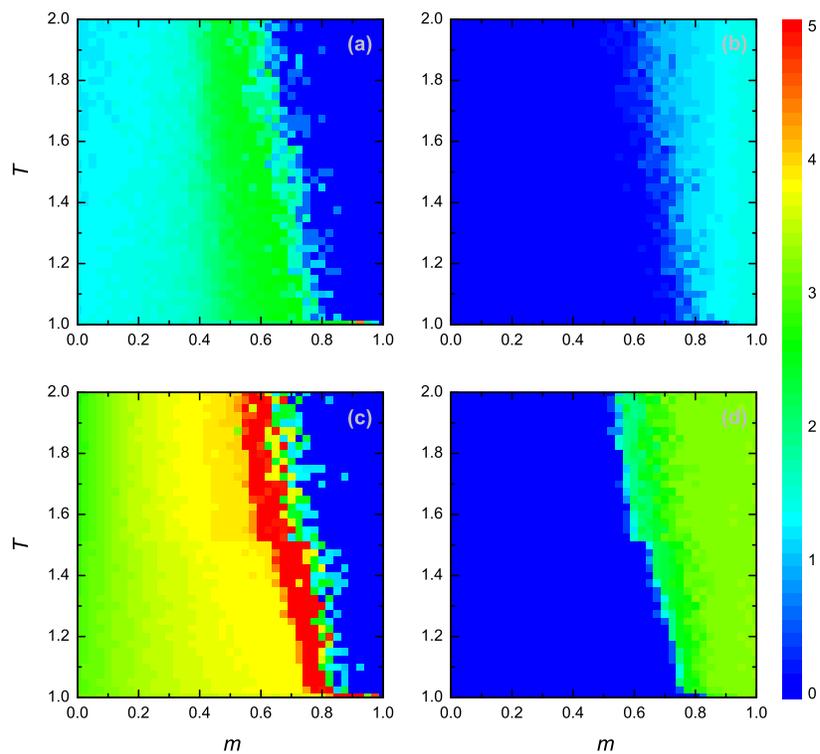}\end{center}
\caption{\textbf{Average degree of cooperators (a,c) and defectors (b,d) for two different population densities [$\rho=0.4$ in (a,b) and $\rho=0.8$ in (c,d)].} Graphical conventions and payoff parameters are the same in Fig.~\ref{fig3}. See main text for further details.}
\label{fig4}
\end{figure}

This interpretation is corroborated by Fig.~\ref{fig4}, which shows
the difference in connectedness between cooperating and defecting
individuals. The regions in parameter space where a high level of
cooperation evolved (Fig.~\ref{fig2}(d)) corresponds to those
regions where cooperators are tightly connected (i.e. where
cooperators have a high degree). As before, connectedness is low in
the low-density situation (where cooperation did not get off the
ground) and much higher (at least for cooperators) in the
high-density situation. For low and intermediate values of $m$, the
connectedness of cooperators is markedly higher than the
connectedness of defectors. The opposite is the case a high values
of $m$.

\begin{figure}
\begin{center}\includegraphics[width=11.015cm]{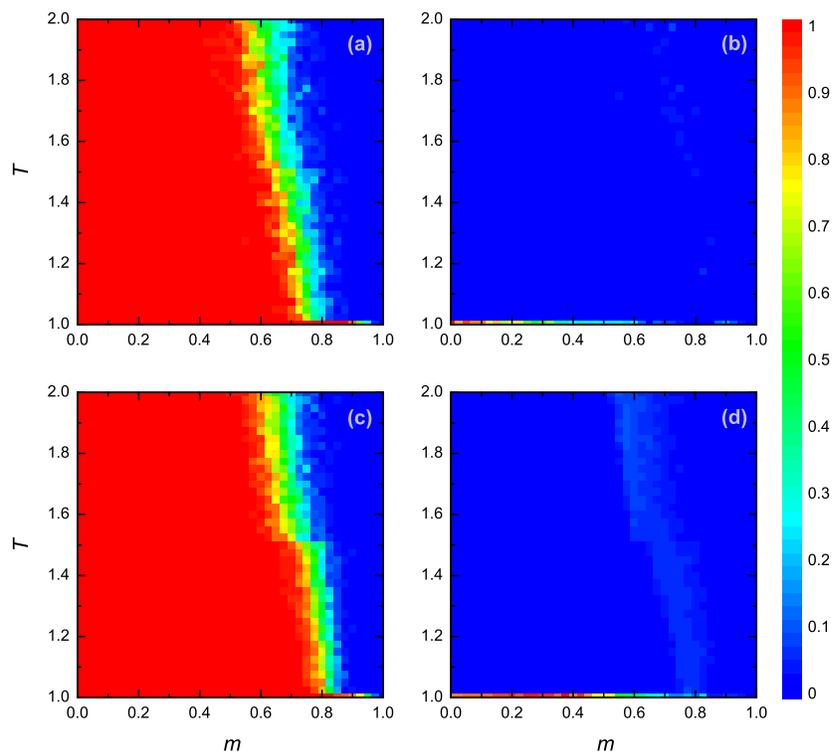}\end{center}
\caption{\textbf{Stationary density of cooperators in the neighborhood of
cooperators (a,c) and defectors (b,d) for two different population densities [$\rho=0.4$ in (a,b) and $\rho=0.8$ in (c,d)].} Graphical conventions and payoff parameters are the same as in Fig.~\ref{fig2}. See main text for further details.}
\label{fig5}
\end{figure}

Finally, we consider the differences in the degree of cooperation
experienced by cooperators and defectors, respectively
(Fig.~\ref{fig5}). Irrespective of population density and other
parameters, defectors always ended up in adverse neighborhoods. In
contrast, cooperators tended to interact only with other cooperators
- at least as long as the fraction of players adopting rewiring was
not too small (i.e. for $m<0.6$). For large values of $m$,
cooperative clusters did not emerge, corresponding to the collapse
of cooperation under these conditions.

Overall, our results confirm the importance of the
formation of cooperative clusters. If population densities are not
too low and if sufficiently many individuals adopt rewiring,
cooperative clusters can emerge even under unfavorable conditions
(e.g. a large value of $T$). Once clusters of cooperators have
formed, selection against defective partners can effectively shield
clusters of cooperators from the invasion of defectors. This is so
because cooperators within the cluster attract interactions with
cooperators at the cluster boundary. In this way the payoffs of
cooperators both inside the cluster and at its fringe are enhanced
and interactions with defectors are avoided. Defectors surrounding
clusters of cooperators have limited opportunities for exploitation,
allowing clusters of cooperators to expand. Thus, clusters uphold
cooperative behavior even if the temptation to defect is large.
Conversely, defectors are unable to claim lasting benefits from
occupying the clusters, simply because they become very weak as soon
as all the neighbors of the defecting cluster become defectors
themselves.

\section*{Discussion}
In line with other studies, we have shown that in a network
environment cooperation in social dilemmas does readily evolve if
players have the opportunity to change their local interaction
structure when surrounded by non-cooperative neighbors. Among the
two reactions to adverse conditions considered, rewiring was clearly
more favorable for the evolution of cooperation than migration. This
was even the case in a model where the costs of rewiring and
emigration were the same (zero), while migration will often be more
costly in natural settings. Interestingly, the highest
degree of cooperation evolved when the player population was
polymorphic in the sense that both types of reaction to adverse
neighborhoods (rewiring and migration) were present in
non-negligible frequencies. We interpret this finding by the
interplay of two factors: while migration induces a certain mixing
of the population due to increased interaction ranges of the
migrating players, rewiring may lead to strongly heterogeneous
interactions networks, which are tightly associated with flushing
cooperative states \cite{szolnoki_epl08}.

In our model, the reaction to adverse conditions was assumed as a
fixed property of each player. Hence, this reaction did not evolve.
Our results suggest that the joint evolution of the strategies in
the cooperation game and the reaction to adverse condition would
lead to a polymorphism in the reaction to adverse neighborhoods. It
remains to be seen whether this is indeed the case.

\section*{Methods}
All simulations were run on a lattice of $3000$ nodes with periodic
boundary. In each simulation, a fraction $\rho$ of the lattice nodes
was occupied by $N=3000 \rho$ players, who initially were
distributed randomly over the lattice. Initially, each player was
connected with all players on adjacent lattice nodes. Accordingly,
the initial degree of each player ranged from $0$ to $4$, with an
average of $4 \rho$.

Throughout a simulation, each player had the fixed status of either
adopting migration or rewiring when confronted with adverse
conditions. This status was assigned to players at random at the
start of a simulation, with $m$ being the fraction of migrants. In
addition, players could at each time be classified as either
cooperators or defectors in the cooperation game, but the strategy
of each player could change during a simulation due to payoff-based
learning. Initially, the strategies $C$ and $D$ were randomly
assigned to the players with equal probability.

To simulate evolution, we employed event-based asynchronous updating
where interactions, rewiring and migration all occurred on the same
time scale. Whenever an ``event'' occurred, a focal player $i$ was
chosen at random. This player had pairwise interactions with all
``neighbors'' (that is, all players connected with $i$), yielding a
sum $P_i$ of all payoffs. For a randomly selected neighbor $j$ of
$i$, the payoff $P_j$ was determined in a similar way. Based on the
payoff difference $P_i-P_j$, the focal player switches to j's
strategy with probability $\{1+\exp[(P_i-P_j)/K]\}^{-1}$. If $K$ is
large, payoff differences do not matter much for the direction of
strategy change, while such differences are decisive in case of a
small value of $K$. Throughout this work we set $K=1$, indicating
that strategies of better performing players are readily, though not
always, adopted.

Following the game interactions and the strategy change
phase, the focal player $i$ reconsiders its interaction structure.
If the focal individual adopts rewiring, it will cut a random tie
with a defecting neighbor with probability proportional to the
number of defectors; subsequently the free link will be reattached
to another player randomly chosen from the entire population. If the
focal player adopts migration, it will migrate to a randomly chosen
empty target site with probability proportional to the number of
defectors in its neighborhood; at the new site, it will establish
new links with all players occupying adjacent sites on the network.

During a full iteration, the above process was repeated $N$ times.
Hence on average each player was in the focal role once per
iteration, and each player had on average once the opportunity to
pass its strategy to one of its neighbors. The process was repeated
until a stationary state was reached, where the distribution of
strategies and the characteristics of neighborhoods did not change
any more. Typically we ran each simulation for $10^6$ steps. For
each parameter combination we ran $100$ replicate simulations. The
results reported are averages over these replicates.


\clearpage

\end{document}